\documentclass[12pt,preprint]{aastex}

\begin{document}

\title{X-ray emission from young brown dwarfs in the Orion Nebula Cluster}

\author{Thomas Preibisch\altaffilmark{1}, 
Mark~J. McCaughrean\altaffilmark{2,3},
Nicolas Grosso\altaffilmark{4}, 
Eric~D. Feigelson\altaffilmark{5},
Ettore Flaccomio\altaffilmark{6},
Konstantin Getman\altaffilmark{5},
Lynne~A. Hillenbrand,\altaffilmark{7},
Gwendolyn Meeus\altaffilmark{3},
Giusi Micela\altaffilmark{6},
Salvatore Sciortino\altaffilmark{6},
Beate Stelzer\altaffilmark{6,8}}

\altaffiltext{1}{Max-Planck-Institut f\"ur Radioastronomie,
Auf dem H\"ugel 69, D-53121 Bonn, Germany}
\altaffiltext{2}{University of Exeter, School of Physics,
Stocker Road, Exeter EX4 4QL, Devon, UK}
\altaffiltext{3}{Astrophysikalisches Institut Potsdam, An der Sternwarte 16,
D-14482 Potsdam, Germany}
\altaffiltext{4}{Laboratoire d'Astrophysique de Grenoble,
Universit{\'e} Joseph-Fourier, F-38041 Grenoble cedex 9, France}
\altaffiltext{5}{Department of Astronomy \& Astrophysics,
Pennsylvania State University, University Park PA 16802}
\altaffiltext{6}{INAF, Osservatorio Astronomico di Palermo G. S. Vaiana, 
Piazza del Parlamento 1, I-90134 Palermo, Italy}
\altaffiltext{7}{Department of Astronomy, California Institute of
Technology, Mail Code 105-24, Pasadena, CA 91125}
\altaffiltext{8}{Dipartimento di Scienze Fisiche ed Astronomiche, Universit\`a 
di Palermo, Piazza del Parlamento 1, I-90134 Palermo, Italy}

\slugcomment{Version: \today}

\begin{abstract}
We use the sensitive X-ray data from the {\it Chandra\/} Orion Ultradeep
Project (COUP) to study the X-ray properties of 34 
spectroscopically-identified brown dwarfs with near-infrared spectral types 
between M6 and M9 in the core of the Orion Nebula Cluster. Nine of the 34 
objects are clearly detected as X-ray sources. 
The apparently low detection rate is in many cases related to the
substantial extinction of these brown dwarfs; 
considering only the BDs with $A_V \leq 5$\,mag, nearly half of the
objects (7 out of 16) are detected in X-rays.
Our 10-day long X-ray 
lightcurves of these objects exhibit strong variability, including numerous 
flares. While one of the objects was only detected during a short flare, a 
statistical analysis of the lightcurves provides evidence for continuous 
(`quiescent') emission in addition to flares for all other objects. Of these, 
the $\sim$\,M9 brown dwarf COUP\,1255 = HC\,212 is one of the coolest known 
objects with a clear detection of quiescent X-ray emission. The X-ray 
properties (spectra, fractional X-ray luminosities, flare rates) of these 
young brown dwarfs are similar to those of the low-mass stars in the ONC, 
and thus there is no evidence for changes in the magnetic activity around 
the stellar/substellar boundary, which lies at $\sim$\,M6 for ONC sources. 
Since the X-ray properties of the young brown dwarfs are also similar to 
those of M6--M9 field stars, the key to the magnetic activity 
in very cool 
objects seems to be the effective temperature, which determines the degree
of ionization in the atmosphere.
\end{abstract}

\keywords{open clusters and associations:  individual (Orion) - stars
- activity - stars:  low-mass, brown dwarfs - stars:  pre-main
sequence - X-rays:  stars}

\section{Introduction \label{intro.sec}}
Brown dwarfs (BDs) are objects with masses below $\sim 0.075\,M_\odot$, the 
stellar/substellar boundary (see Basri 2000 for a review).
In contrast to stars, which derive their luminosity from hydrogen fusion, 
BDs never reach sufficiently high core temperatures to start hydrogen burning, 
though brief episodes of deuterium and lithium burning occur early in their 
evolution.  With no sustainable internal fusion energy source, BDs 
continuously cool down and dim with time.  During the first few Myr of their 
evolution, BDs are thus warmer and orders of magnitude brighter than at 
older ages: for example, between the ages of 1\,Myr and 5\,Gyr, a 
$0.03\,M_\odot$ BD cools from $T_{\rm eff} = 2660$\,K down to 
$T_{\rm eff} = 600$\,K, and its luminosity drops by four orders of magnitude 
from $\log\left(L_{\ast}/L_{\odot}\right) = -2.1$ to 
$\log\left(L_{\ast}/L_{\odot}\right) = -6.1$ (Baraffe et al.\ 1998). Young BDs 
can thus be readily detected at larger distances much more easily than 
older BDs and as a consequence, numerous young BDs have recently been 
discovered in several nearby star-forming regions. The largest population 
is found in the Orion Nebula Cluster (ONC; McCaughrean et al.\ 1995; 
Hillenbrand \& Carpenter 2000; Muench et al.\ 2002). 

Even when young, BDs are relatively cool and dim objects, and one would
not intuitively expect them to emit high energy radiation.  Also, as BDs 
have a fully convective internal structure, they cannot possess a solar-like 
$\alpha$--$\Omega$ dynamo, which is thought to be the energy source of X-ray 
activity in late-type stars.  Nevertheless, some brown dwarfs have been
detected as X-ray sources ($\S 2$). The nature of the X-ray 
emission from BDs (and similarly from fully convective very-low mass stars) 
and the origin of their activity is still not well understood.

In this paper, we focus on a relatively small sample of 
spectroscopically-confirmed BDs in the ONC classified by Slesnick, Hillenbrand, 
\& Carpenter (2004; henceforth SHC04) and their X-ray properties as measured 
in \dataset[ADS/Sa.CXO#obs/COUP]{the {\it Chandra\/} Orion 
Ultradeep Project (COUP)}. A significantly larger
sample of candidate BDs has been identified in the ONC by several authors 
based on broad-band optical and near-infrared photometry, but their status 
as true BDs remains unconfirmed by spectroscopy. In a subsequent paper, we
will present an analysis of the X-ray properties of the larger sample
of very low-mass stars and BD candidates in the region as detected in 
deep near-infrared imaging photometry obtained with the ESO Very Large 
Telescope and characterised in terms of their photometric properties 
alone (McCaughrean et al.\ 2005, in preparation).

\section{Previous X-ray detections of spectroscopically confirmed BDs}
The first (and at that time unrecognized) detection of X-ray emission from 
a BD was made as early as 1991, when {\it ROSAT\/} obtained a deep X-ray image 
of the Chamaeleon star-forming region.  One of the weak X-ray sources was 
identified with a faint ($V \sim 21$ mag) point source, for which no further 
information was available at the time. Several  years later, after the first 
confirmed BDs were announced in 1995, Neuh\"auser \& Comer\'on (1998) presented 
an optical spectrum of this object, Cha\,H$\alpha$\,1, and derived a spectral 
type of M7.5, from which they inferred a substellar mass of 
$\sim 0.05\,M_\odot$. This episode serves to demonstrate that the main 
obstacle in the investigation of X-ray emission from BDs is often not their 
detection as faint X-ray sources, but rather the lack of optical/near-infrared 
spectroscopy necessary for reliable mass estimates.

Although some further BD detections were made with {\it ROSAT\/} (e.g., 
Comer\'on et al.\ 2000), the X-ray fluxes were only marginally above the 
detection limits and the $\sim 15$\arcsec{} spatial resolution of the
satellite often led to identification difficulties. The advent of the 
new X-ray observatories {\it XMM-Newton\/} and {\it Chandra\/} boosted the 
effort with their greatly increased sensitivity and, in the case of
{\it Chandra\/}, superb spatial resolution ($\sim 1$\arcsec). For 
example, X-rays were detected from two BDs in the $\rho$\,Oph star-forming 
region by both satellites and showed strong long-term variability (Imanishi 
et al.\ 2001; Ozawa et al.\ 2005).  A {\it Chandra\/} observation of the young
cluster IC\,348 provided X-ray detections of 7 very low-mass objects, 4 of 
which are spectroscopically-confirmed BDs (Preibisch \& Zinnecker 2001, 2002).  
Tsuboi et al.\ (2003) resolved X-rays from the $\sim 10$\,Myr old M8.5--9 
BD TWA-5B separated by 2\arcsec{} from its primary, TWA-5A, in the nearby 
TW\,Hya association, while Gizis \& Bharat (2004) saw no emission from
2MASS~J1207334$-$393254, another M8 BD in the same association. The former 
was detected in quiescence at a level of $\log L_{\rm X} = 27.6$~erg/sec and
$\log \left(L_{\rm X}/L_{\rm bol}\right) = -3.4$ with an unusually soft spectrum, 
while for the latter, upper limits of $\log L_{\rm X} < 26.1$~erg/sec and 
$\log \left(L_{\rm X}/L_{\rm bol}\right) < -4.8$ were determined. Finally, in the 
Chamaeleon~I cloud, Feigelson \& Lawson (2004) reported X-rays from three 
objects around the substellar limit in the northern molecular core using 
{\it Chandra\/}, while Stelzer et al.\ (2004) detected two 
spectroscopically-confirmed BDs and several BD candidates in the southern 
core using {\it XMM-Newton}.

To date, just two older field BDs have been detected in X-rays. The first of 
these is the nearby ($d = 5$\,pc) M9 dwarf LP\,944-20, with an estimated mass 
of $\sim 0.06\,M_\odot$ and age of $\sim 600$\,Myr. Rutledge et al.\ (2000) 
discovered an X-ray flare from LP\,944-20 during a {\it Chandra\/} observation, 
but detected no quiescent emission. At the flare peak, the X-ray luminosity 
was $\log L_{\rm X} = 26.1$~erg/sec and $\log\left(L_{\rm X}/L_{\rm bol}\right) \sim -3.7$.  
A subsequent deep {\it XMM-Newton\/} observation by Mart\'{\i}n \& Bouy (2002) 
provided a very sensitive upper limit of $\log L_{\rm X} < 23.5$~erg/sec and 
$\log\left(L_{\rm X}/L_{\rm bol}\right) < -6.3$ for possible quiescent 
emission. A powerful X-ray flare and probable quiescent 
emission\footnote{The quiescent emission was only seen after, but not before 
the flare, and thus may be an afterglow of the flare.}, was detected from 
the second field source, the M8.5$+$M9 binary Gl\,569\,Ba,b, which orbits 
the nearby ($d = 10$\,pc), $\sim$\,300--800\,Myr old M2 star Gl\,569\,A with 
5\arcsec{} separation (Stelzer 2004). A dynamical mass determination for the 
components in the $0.1$\arcsec{} binary Ba,b gives 
$M_{\rm Ba} = 0.055$--$0.087\,M_\odot$ and 
$M_{\rm Bb} = 0.034$--$0.070\,M_\odot$ (99\% confidence intervals; Zapatero 
Osorio et al.\ 2004); the lower-mass component at least is the first
model-independently confirmed substellar object.

\section{X-ray emitting brown dwarfs in the ONC}
\subsection{Previous X-ray results on very-low mass objects in the ONC}
Prior to COUP, the ONC had been observed with both imaging instruments, ACIS 
and HRC, aboard {\it Chandra\/}. The results of two ACIS-I observations with 
a combined exposure time of 23 hours were reported in Garmire et al.\ (2000) 
and Feigelson et al.\ (2002ab, 2003).  These ACIS observations revealed
X-ray detections of about 30 very low-mass objects (Feigelson et al.\ 2002a);
for most of these objects, however, no optical/infrared spectra were available
and it was therefore unclear whether they were BDs or low-mass stars. Several 
of the very-low mass objects showed X-ray flares, which appeared to be similar 
in frequency and morphology to the flares on low-mass stars.
% 1/4 have lxlbol between -3 and -4
Feigelson et al.\ (2002a) concluded that the candidate very-low mass objects 
had X-ray properties similar to those of low-mass stars and that magnetic 
activity appears to decline as the very-low mass objects evolve.
Flaccomio et al.\ (2003a,b) presented an analysis of their 17.5\,hr HRC-I
observation of the ONC\@.  From a `composite source analysis' of a sample 
of very-low mass objects (one X-ray detected object and 14 upper limits), they
concluded that the BDs seem to follow the same $L_{\rm X}\leftrightarrow M$ 
relationship as low-mass stars.

\subsection{The ONC brown dwarf sample} \label{sample.sec}
The presence of an extensive substellar population in the ONC has been 
deduced in a series of studies (e.g., McCaughrean et al.\ 1995; Hillenbrand 
\& Carpenter 2000; Luhman et al.\ 2000; Lucas \& Roche 2000; Muench et al.\ 
2002). These studies, however, were generally based on photometry alone, 
which can lead to ambiguities when trying to establish the membership and 
mass of a specific source, as opposed to the properties of the ensemble 
population. To derive more definitive properties for a subset of sources, 
SHC04 recently presented a spectroscopic study of candidate BD members in the 
central $5.1\times 5.1$ arcmin ($\sim 0.7\times 0.7$\,pc) part of the ONC\@. 
Using near-infrared and optical spectra, they derived spectral types and basic 
parameters, such as bolometric luminosity and extinction, for about 100 faint 
objects, from which they then constructed an H-R diagram. Masses were inferred 
using the D'Antona \& Mazzitelli (1997) evolutionary tracks, leading 
to the identification of 34 objects with masses nominally below 
$0.075\,M_\odot$.

It should be noted that the majority of the SHC04 spectral classifications 
were based on near-infrared spectra, as less than half of the sources in their 
sample had corresponding optical spectra. Conversely, a small handful of 
sources had {\em only\/} optical spectra which were then used for their 
classification. At issue here is the well-known fact that near-infrared 
classification tends to yield systematically later spectral types for young 
brown dwarfs than given by optical spectral classification (see, e.g., Luhman 
\& Rieke 1999; Luhman et al.\ 2003). Typically, there is a shift to later 
types by $\sim$\,1 subclass for M6--M7 sources, and by $\sim$\,1--2 subclasses 
for types at M8 and later. This systematic shift can 
be seen in the ONC sources of SHC04 for which both near-infrared and optical 
spectra were available (their Tables~1 and~2). 

Another important point to note is that these very young objects should have 
relatively low surface gravities which can lead to significant changes in the 
depth of some of the traditional classification indices; e.g., those based on 
the near-infrared water absorption features, in turn causing errors in the 
resulting spectral types, effective temperatures, and placement in the H-R 
diagram (see, e.g., McGovern et al.\ 2004; Gorlova et al.\ 2003; Lucas et al.\ 
2001; Wilking et al.\ 2004). SHC04 addressed the spectral typing issue by 
measuring standards drawn from a range of high surface gravity main sequence 
stars, low surface gravity sources from relatively young clusters,
and field giants. However, when converting from spectral types 
to effective temperatures, they used a high gravity dwarf temperature scale,
pointing out that young pre-main sequence objects appear observationally 
closer to dwarfs than giants or subgiants, and also that no accurate 
temperature or bolometric correction scales for pre-main sequence stars 
are available to date.

\begin{figure}[p]
\plotone{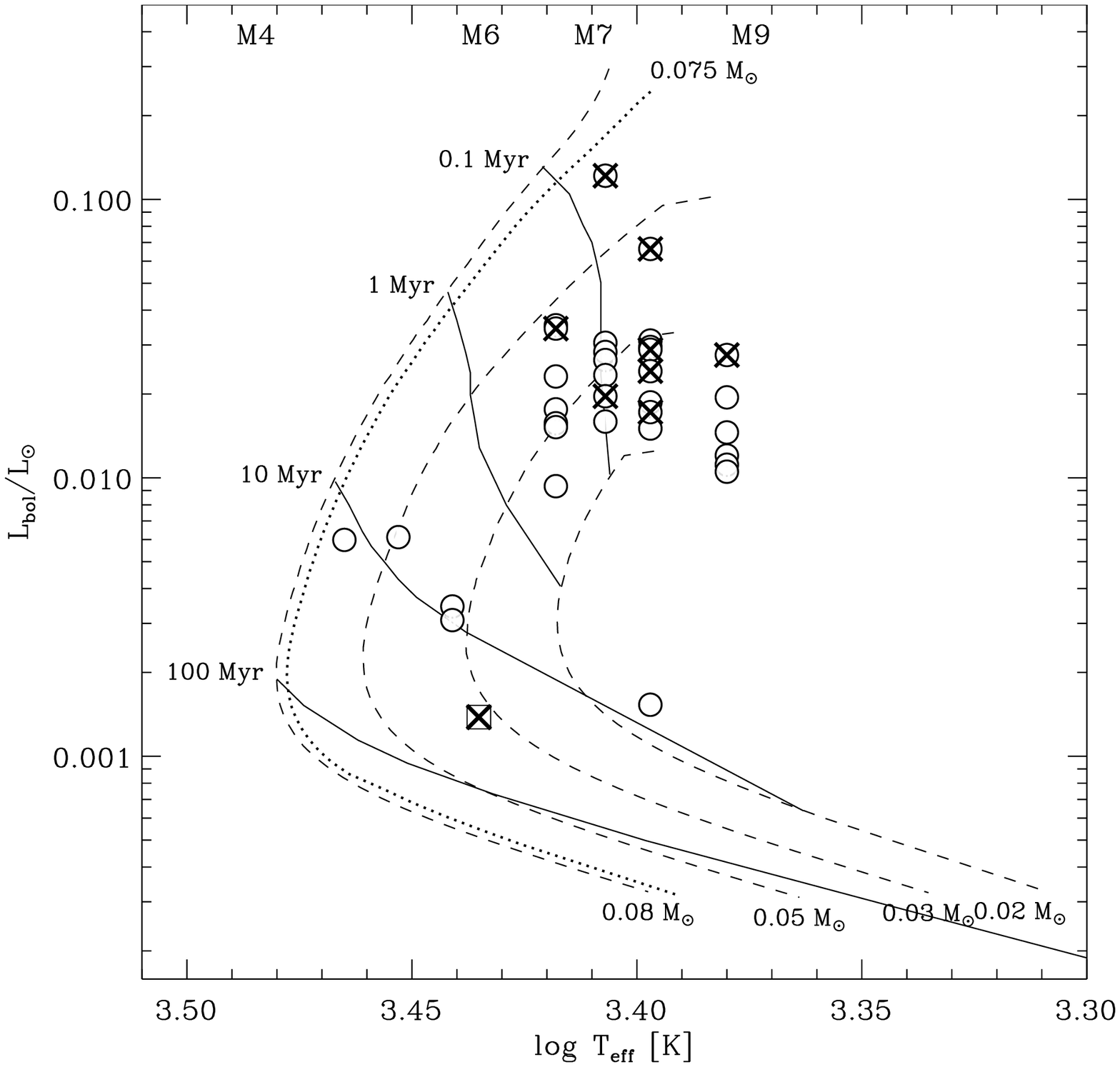}
\caption{
H-R diagram for the spectroscopically-confirmed ONC BDs (circles) from SHC04.
HC\,722, which is the only object for which the spectral analysis suggested 
a high surface gravity (rather than the low gravity expected for very young
BDs), is plotted as a square.
Objects detected as X-ray sources in the COUP data are marked by crosses.
The evolutionary tracks (dotted lines) and isochrones (solid lines) are
from D'Antona \& Mazzitelli (1997).
\label{hrd.fig}}
\end{figure}

By contrast, other authors (e.g., White et al.\ 1999; Luhman et al.\ 2003) 
have used a spectral type to effective temperature conversion based on optical 
spectral types and fitting the isochrones of Baraffe et al.\ (1998) for young 
brown dwarfs. They suggest that for a given optical M~spectral type, young 
objects have effective temperatures $\sim$100--200\,K warmer than field 
dwarfs. This compounds the near-infrared/optical shift 
in spectral type. For example, consider a young source with an optical spectral 
type of M7 and, for the sake of argument, an M8 spectral type derived from a 
near-infrared spectrum. The field dwarf temperature scale adopted by SHC04 
gives $\sim$2500\,K for spectral type M8, and $\sim 2620$ K for M7.
The pre-main sequence temperature scales of White 
et al.\ (1999) and Luhman et al.\ (2003) give $\sim$\,2850\,K as the effective 
temperature for the optical spectral type M7. Thus, in this hypothetical 
example, the near-infrared classification and field dwarf temperature scale would yield 
an effective temperature of 2500\,K for the source, where the optical 
classification and PMS temperature scale would give 2850\,K. 
From Fig.~\ref{hrd.fig}, it can be seen that such a temperature
shift would result in an upwards revision of the mass of the source by a
factor of 2--3 using the D'Antona \& Mazzitelli (1997) tracks.

A further problem is that there is substantial detailed disagreement between 
various pre-main sequence evolutionary tracks covering the stellar/substellar 
transition at early ages (e.g., those of D'Antona \& Mazzitelli 1997 as used 
by SHC04 versus the models of Baraffe et al.\ 1998 or Siess et al.\ 2000 used 
elsewhere; see, for example, the analyses of Luhman 1999 and Hillenbrand \&
White 2004). In particular, Hillenbrand \& White (2004) concluded that none
of the presently available sets of evolutionary models provide a fully
satisfactory match between masses derived on the basis of pre-main sequence
tracks and those measured dynamically: there is a general trend that they
underpredict mass by 10--30\% in the range 1.2--0.3\,M$_\odot$, 
the lowest masses presently testable in this manner.

These issues are too involved to be analysed in detail here. Furthermore,
it is not practical to rederive the spectral types, effective temperatures,
and masses for the SHC04 sample here. For example, roughly 75\% of the 34 
sources classified as BDs by SHC04 do not have optical spectra and thus
near-infrared spectral types must be used. As a consequence, it is possible 
that some of the 34 sources that SHC04 spectroscopically classified as BDs 
do in fact have stellar masses, but we shall assume for present purposes that 
all of them are indeed true BDs. 
We 
focus on the nominal BD sources exclusively, omitting the objects 
in the SHC04 sample with masses nominally in the stellar regime.
These, and the numerous additional BD candidates seen in
 near-infrared images, will be treated in a separate paper.

Finally, we note that the latest spectral type found 
by SHC04 was L0, for the source HC\,722. As this source was also detected in 
the COUP data during a flare (COUP\,344) as discussed below, it would have 
represented the latest spectral 
type source seen in X-rays to date. However, a careful re-assessment of the 
SHC04 optical spectrum for HC\,722 (Slesnick et al.\ 2005) 
has shown that its spectral type is in fact significantly earlier, at around 
M6--6.5. We adopt this revised spectral type in the remainder of our analysis, 
along with the corresponding new physical parameters $\log T_{\rm eff} = 3.435$,
$\log \left(L_{\rm bol}/L_\odot\right) = -2.88 $, and a mass of 0.035\,M$_\odot$. It is also
important to note that HC\,722 was the only COUP-detected BD candidate
showing indications of a high surface gravity, suggesting that it may
in fact be a foreground field star. This possibility is discussed in more
detail in \S\ref{hc722.sec}.

\subsection{The COUP observation}
The COUP observation of the ONC is the longest and deepest
X-ray observation ever made of a young stellar cluster, providing a rich and 
unique dataset for a wide range of science studies.  Full observational 
details, a complete description of the data analysis, and the definitions of 
the derived X-ray quantities can be found in Getman et al.\ (2005a).  Briefly,
the total exposure time of the COUP image with ACIS-I on {\it Chandra\/}
was 838\,100~sec (232.8 hours or 9.7 days) with a single 17$\times$17 arcmin
FOV pointing centered near the Trapezium stars. A total of 1616 individual 
X-ray sources were found in the COUP image, and the superb PSF and the high 
accuracy of the aspect solution allowed a clear and unambiguous identification 
of $\simeq 1400$ X-ray sources with near-infrared or optical counterparts, with
median offsets of just 0.15\arcsec{} and 0.24\arcsec, respectively 
(Getman et al.\ 2005b).  The X-ray 
luminosities of the sources were determined in the spectral fitting analysis; 
integrating the best-fit model source flux over the [$0.5 - 8.0$]\,keV band 
yielded the intrinsic (extinction-corrected) X-ray luminosity 
($L_{\rm X} := L_{\rm t,c}$  in
the nomenclature of Getman et al.\ 2005a).
The detection limit of the COUP data is 
$\log L_{\rm X} = 27.3$~erg/sec for lightly absorbed 
sources. Given the typical bolometric luminosities of young BDs in the ONC 
of $10^{-1}$--$10^{-3}\,L_\odot$, we are thus able to probe the X-ray activity 
of these objects down to levels of 
$L_{\rm X}/L_{\rm bol} \sim 10^{-5}$--$10^{-3}$.

\subsection{X-ray detected BDs in the COUP}

Nine of the 34 spectroscopic BDs (objects with mass estimates 
$<0.075\,M_\odot$) in the SHC04 sample are detected as X-ray sources in the 
COUP data set and the observed X-ray properties of these objects are discussed 
in detail below. A summary of their 
optical/near-infrared properties is given in Table~\ref{NIRprop.tab}, while
their X-ray properties are listed in Table~\ref{Xrayprop.tab}. For the 25 
spectroscopic BDs not detected as X-ray sources, upper limits to their 
X-ray count rates were derived as described in the next subsection. In 
Fig.~\ref{hrd.fig}, we show an H-R diagram with the all the spectroscopic
BDs in the SHC04 sample. It is clear that the COUP detection likelihood is 
a strong function of the bolometric luminosity; this reflects the correlation 
between X-ray and bolometric luminosity (see below). Another factor limiting 
our ability to detect X-ray emission from the BDs is their extinction, as 
will be discussed in \S\ref{det_frac.sec}.

Two aspects should be noted with respect to the derived X-ray luminosities
of the BDs.  First, it should be kept in mind that since the X-ray 
luminosities were determined from the fits to the temporally averaged spectra 
obtained over the full $\sim$\,10 day COUP exposure, the values for the X-ray 
luminosities are also temporal averages over the same period. During the flares 
seen in the individual lightcurves, the luminosities can be considerably 
higher. In \S\ref{lightcurves.sec}, we also determine the `characteristic' 
values of the X-ray luminosities, which can be thought of as the typical  
or quiescent level of X-ray energy output of these sources excluding 
flares. The only exception is COUP\,344 (HC\,722), which was only detected 
during a flare; for this object we estimated the flare luminosity and an 
upper limit to the X-ray luminosity outside the flare period (see 
\S\ref{hc722.sec}).

Second, since the number of source photons per spectrum is rather small 
($< 100$), the reliability of the extinction-corrected values for the X-ray 
luminosities, $L_{\rm t,c}$, is not immediately clear. We have therefore 
compared the values for the absorbing hydrogen column density found in the 
spectral fits to those expected from the visual extinction according to the 
relation $N_{\rm H} = A_V \times 1.6 \times 10^{21}\,\rm cm^{-2}$ (Vuong et 
al.\ 2003), where $A_V$ is that determined by SHC04 from
their optical and near-infrared spectra. In all cases, the $N_{\rm H}$ values 
found in the X-ray spectral fits are either consistent with or somewhat lower 
than the estimate based on $A_V$. This implies that our X-ray luminosities 
are reliable and not affected by potential problems due to overestimates of 
the absorbing hydrogen column density in the X-ray spectral fits.

\subsection{Determination of upper limits for the undetected BDs}
While the COUP data are generally sensitive to stars with luminosities above 
$\log L_{\rm X} \rm{(lim)} \simeq 27.0-27.5$~erg/sec for lightly absorbed 
stars (Getman et al.\ 2005a), individual X-ray upper limits can be obtained 
for undetected ONC members when an estimate of the absorption is available.
In Table~\ref{Slesnick_UL.tab}, we give COUP X-ray upper limits for those 
25 BDs from SHC04 that were undetected in COUP, where
these upper limits were obtained from the COUP image as follows.  Photons 
were extracted at the position of each star using the {\it ACIS Extract\/} 
procedure described in Getman et al.\ (2005a).  This procedure subtracts a 
local background from a region containing $\simeq 90$\% of the expected 
events based on the shape of the {\it Chandra\/} point spread function at 
that location in the field.  Column~2 of Table~\ref{Slesnick_UL.tab} gives 
the Poissonian 68\% upper confidence level of source counts based on the 
extracted and background counts, but is truncated at 4 counts as a realistic 
lower limit. The table then provides COUP effective exposure times, which 
include telescope vignetting and other instrumental effects, and visual 
absorption estimates from the spectroscopy by SHC04. 
Conversion factors from ACIS-I count rates to X-ray fluxes are given for the 
observed emission (Column~5) and intrinsic emission corrected for absorption 
(Column~6). We have used the PIMMS software developed by the NASA High Energy 
Astrophysics Science Archive Science Center, with an assumed intrinsic source 
spectrum of a 1\,keV thermal plasma with a metal abundance of 0.4 times solar,
typical for faint ONC X-ray sources (Getman et al.\ 2005a). The final 
table columns give the resulting observed and intrinsic X-ray luminosity 
limits assuming a distance of 450\,pc to the ONC\@. 

\subsection{Detection fraction\label{det_frac.sec}}
\begin{figure}[p]
\plotone{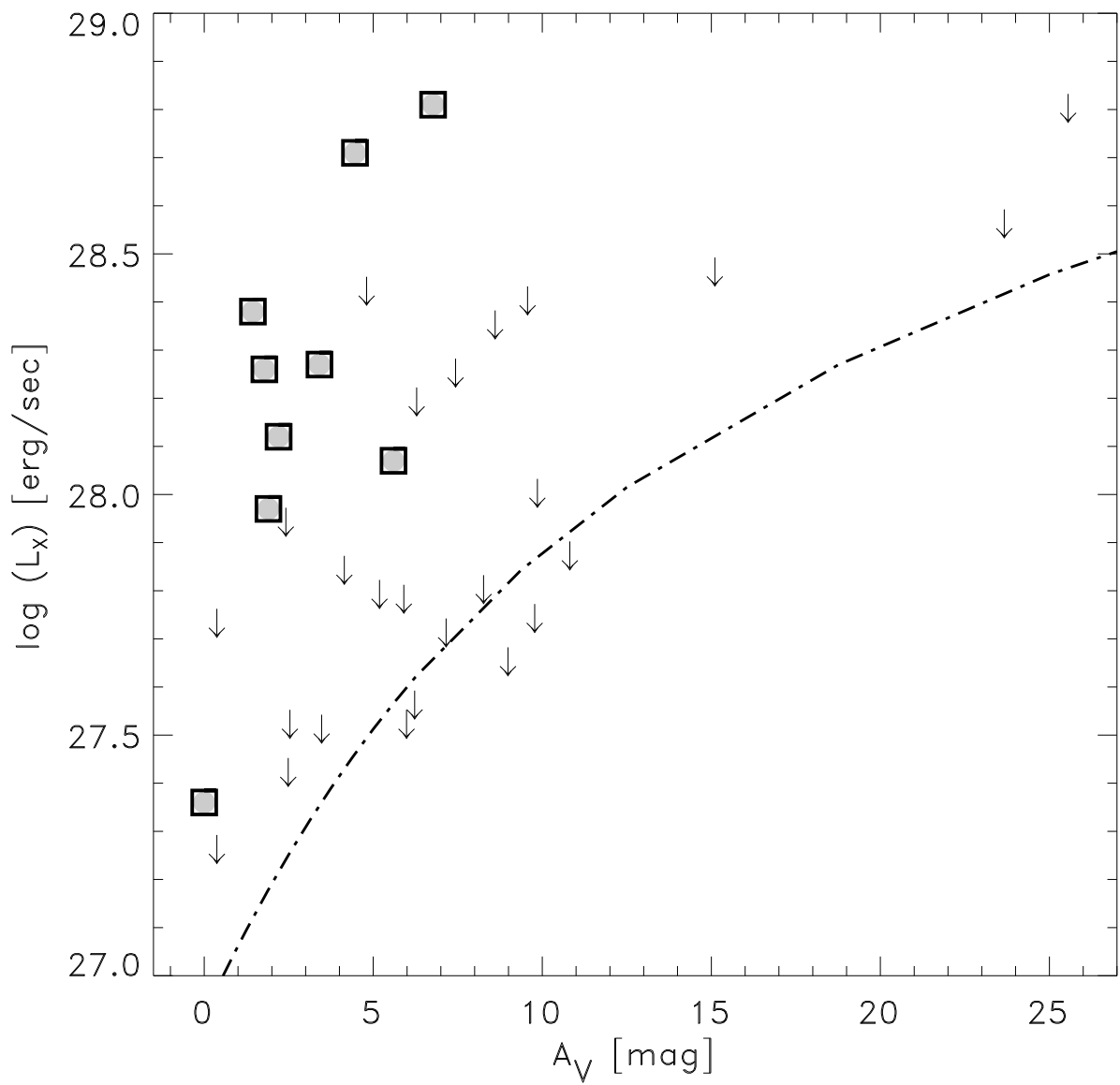}
\caption{X-ray luminosities (extinction-corrected average values during the
period of the COUP observation) versus optical extinction of the BDs from
SHC04. The grey filled boxes show the X-ray luminosities for the detected BDs,
the arrows mark the upper limits for undetected BDs. The
dot-dash line shows the theoretical COUP sensitivity limit for detections
with 5 source counts.
\label{lx_av.fig}}
\end{figure}

Why are only 9 of the 34 (26\%) spectroscopically confirmed BDs detected in
the COUP data? This low detection fraction is related not only to the 
intrinsic faintness of the BDs, but also to their extinction, since the X-ray 
detection limit increases as a function of the extinction. Many of the BDs in 
the SHC04 sample suffer from substantial extinction, up to $A_V \sim 25$\,mag. 
Fig.~\ref{lx_av.fig} shows that X-ray emission is preferentially detected from 
the BDs with relatively low extinction: the detection fraction is 7/16 = 44\% 
for the BDs with $A_V \leq 5$\,mag, but only 2/18 = 11\% for the BDs with 
$A_V > 5$\,mag.  

\section{Temporal and spectral X-ray characteristics of the BDs}
Due to the intrinsic X-ray faintness of substellar objects, most previous 
X-ray observations of BDs have yielded only very small numbers of photons, 
often too few to allow a reasonable spectral or temporal analysis.  Even the 
extraordinary deep exposure of the COUP dataset yields only moderate numbers 
($\lesssim 100$) of source photons for the X-ray detected BDs in the ONC\@. 
Nevertheless, these are enough to allow us to retrieve important information 
from the X-ray data.

\subsection{Lightcurve analysis\label{lightcurves.sec}}
\begin{figure}[p]
\plotone{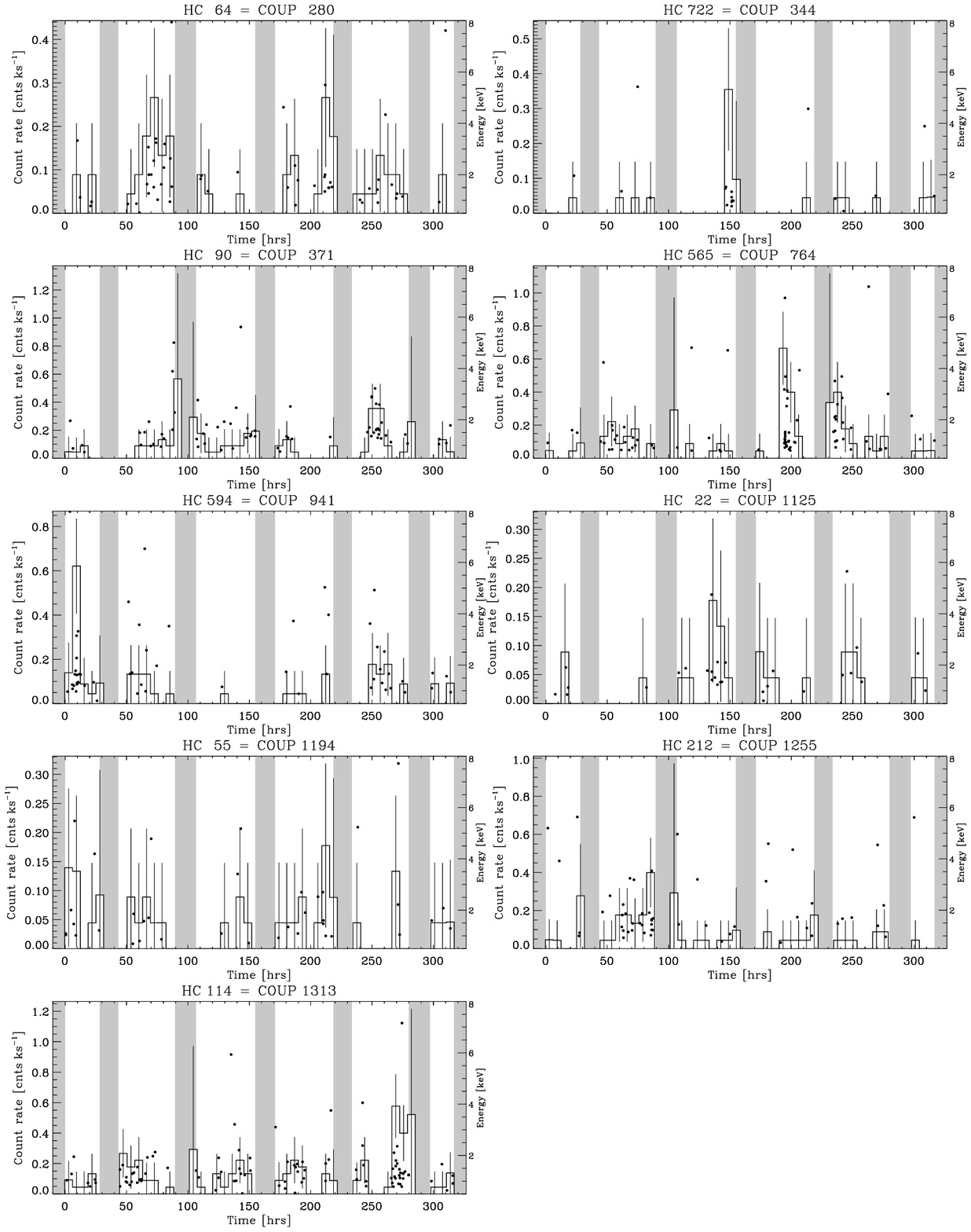}
\caption{Histogram representation of the lightcurves of the 9 X-ray detected
spectroscopic BDs from SHC04 in the $[0.5 - 8]$\,keV band. The solid dots mark
the energies and arrival times of the individual photons in each source area.
The grey stripes mark the time periods when {\it Chandra\/} was not observing.
\label{light_mos.fig}}
\end{figure}

Fig.~\ref{light_mos.fig} shows the lightcurves of the X-ray detected BDs. 
All objects show evidence for rather strong variability; in most cases, 
flare-like bursts are seen. In order to characterize the variability in an 
objective way, the Maximum Likelihood Blocks (MLB) algorithm has been used, 
which segments the lightcurve into a contiguous sequence of constant count 
rates (for a full description of the method, see Wolk et al.\
2005); this makes it possible to determine the number of flares and 
the characteristic level of X-ray emission objectively.  The MLB 
algorithm is similar to the Bayesian Block analysis, the application of which 
to the COUP data is described in Getman et al.\ (2005a), but attempts to 
overcome one of the limitations of the latter technique, namely that it is 
able to segment a light curve into only two segments at a time. This appears
to be a drawback in the search for faint impulsive events (e.g., flares), as 
a two-segment representation of the light curve in which one segment includes 
the event, might not be statistically significant, preventing the segmentation 
process from starting.  
Here we are mainly interested in the number of flares 
and in the characteristic count rate, which can be seen as an estimate of the 
`quiescent' level of X-ray emission, as opposed to the average count rate,
which simply includes all X-ray counts measured, whether in or out of a flare.  
We note that the meaning of `quiescent' 
X-ray emission is not clear; for example, apparently quiescent emission may, 
in reality, just be a superposition of numerous unresolved flares. 
Nevertheless, the 
characteristic level can be taken to describe the `usual' X-ray output of the 
source, outside periods of strong flares.  The results of the MLB analysis
are listed in Table~\ref{countrates.tab}.

We first consider the flares in the X-ray lightcurves of the BDs. The total 
number of flares identified by the MLB algorithm in the 9 lightcurves is 13,
which corresponds to a flare rate of about one flare per object per 180 hours,
consistent with the flare rate of one per 200 hours derived for a sample of
28 solar-like stars in COUP by Wolk et al.\ (2005). In this
regard at least, the temporal characteristics of the X-ray emission from the 
BDs therefore appear to be quite similar to those of low-mass stars in the
ONC.

The second important result from the MLB analysis of the lightcurves is that 
a characteristic level could be definitively established for all spectroscopic 
BDs with the exceptions of COUP\,344 (HC\,722), which was detected only during 
a flare and which will be discussed in more detail in \S\ref{hc722.sec}, 
and COUP\,941 (HC\,594), where it was measured at less than $3\sigma$ 
significance. 
For the other BDs, however, the characteristic level is clearly established 
and demonstrates that these objects would have been detected as X-ray sources 
even without the flares. It also implies that the sources do tend to produce 
X-ray emission in a more continuous manner than just during occasional large 
flares. The detection of apparently `quiescent' emission from these BDs is 
important with respect to the origin of the emission, as discussed further
in \S\ref{origin.sec}.

The characteristic count rates are also used to compute an estimate for the 
characteristic X-ray luminosity by multiplying the temporally averaged X-ray 
luminosity, as determined from the spectral analysis, by the ratio of the 
characteristic countrate to the mean countrate over the COUP exposure. We 
note that this scaling of the luminosities is not fully self-consistent, 
because the X-ray spectral parameters (e.g., plasma temperatures) may change 
as a function of the emission level, whereas the X-ray luminosity was 
determined from the fits to the full, temporally-averaged spectra. A fully 
self-consistent determination of the characteristic X-ray luminosities would 
require time-resolved spectroscopic analysis, but due to the low number of 
detected source counts per BD, this is not possible.

Comparison with the pre-COUP X-ray observations of the ONC provides an 
opportunity to look for long-term variability of the X-ray sources. Four of 
the X-ray detected BDs (COUP\,280, 371, 1125, and 1313) were detected as 
X-ray sources in the previous 23\,hr {\it Chandra\/} ACIS-I observation 
discussed by Feigelson et al.\ (2002a). The 23\,hr observation lightcurves of 
all four sources were classified as `constant' and their reported X-ray 
luminosities are generally quite consistent with those derived from the
COUP data to within a factor of $\sim$\,2--3: the only exception is COUP\,371 
(HC\,90), for which a $\sim 6$ times higher luminosity was derived from the 
23\,hr observation than found here from the COUP data. 

Most of the COUP 
detected BDs that remained undetected in the 23\,hr observation yielded less 
than 50 counts in the 233\,hr COUP dataset,
giving a consistent low level of X-ray emission.  
Conversely, one source
in the earlier dataset, CXOONC 053518.2-052346, 
coincides with an SHC04 BD, HC\,221 (spectral type M7.5), which was undetected
in COUP.
 Fourteen counts were detected giving
 $\log L_{\rm t} \sim 28.5$ in the earlier observation,
and no evidence for variability was reported. 
Using PIMMS, we estimate 
an extinction-corrected X-ray luminosity of $\log L_{\rm X} \sim 29.0$~erg/sec, 
yielding a fractional X-ray luminosity of 
$\log\left(L_{\rm X}/L_{\rm bol}\right) \sim -3.1$ for HC\,221.
The upper limit derived from the COUP data is a factor of about 4 lower than 
the X-ray luminosity measured in the 23~hr observation.

These comparisons suggest that the level of activity in most of the BDs did
not change dramatically over the several years between October 1999/April 
2000 when the 23\,hr observation was obtained and January 2003 when the COUP 
observations took place. There is evidence for changes by more than a factor 
of $\sim 4$ in just two sources.

\subsection{The X-ray flare on COUP\,344 (HC\,722)}
\label{hc722.sec}
The object \object[COUP 0344]{COUP\,344} (HC\,722) deserves special attention since it was only 
detected during an X-ray flare. As noted above, HC\,722 is the only 
COUP-detected object in the SHC04 sample for which a high gravity was 
found from the spectral analysis, rather than a low gravity as seen for 
almost all other young members of the ONC\@. This finding and the extinction 
estimate of $A_V = 0$~mag tend to imply that COUP\,344 is not a member of the 
ONC, but that it is more likely to be an older, foreground field source. 
Keeping in mind the revised M6--6.5 spectral type and $T_{\rm eff} = 2720$\,K
for COUP\,344 as mentioned above, and assuming that 
COUP\,344 is 1\,Gyr rather than 1\,Myr old, 
it would have a mass of $\sim 0.095$~M$_\odot$ (i.e.\ it would be a star and
not a BD) and lie at a distance
$\sim 330$~pc with time-averaged X-ray luminosity $\log L_X \sim 27.3$.

Whether or not the object is a member of the ONC, the time-averaged
 X-ray luminosity is a poor representation of its flaring behavior.
 Eight of the 18 counts detected from COUP 344 outside the 9 hour flare
 are consistent with the measured background rate.  Adopting an upper
 limit of 5 true source photons during this period gives a limit to
 the quiescent emission of $\log L_{\rm X, q} < 27.1$~erg/sec with
 $\log\left(L_{\rm X, q}/L_{\rm bol}\right) < -3.6$. During the flare, 8 counts arrived
 of which none are likely to be background.  The flare luminosity
 averaged over this 9 hr period is $L_{\rm X,f} \simeq 28.7$~erg/sec
 with $\log\left(L_{\rm X, f}/L_{\rm bol}\right) \simeq -2.0$.

\subsection{X-ray spectra and plasma temperature \label{spectra.sec}}

\begin{figure}[p]
\plotone{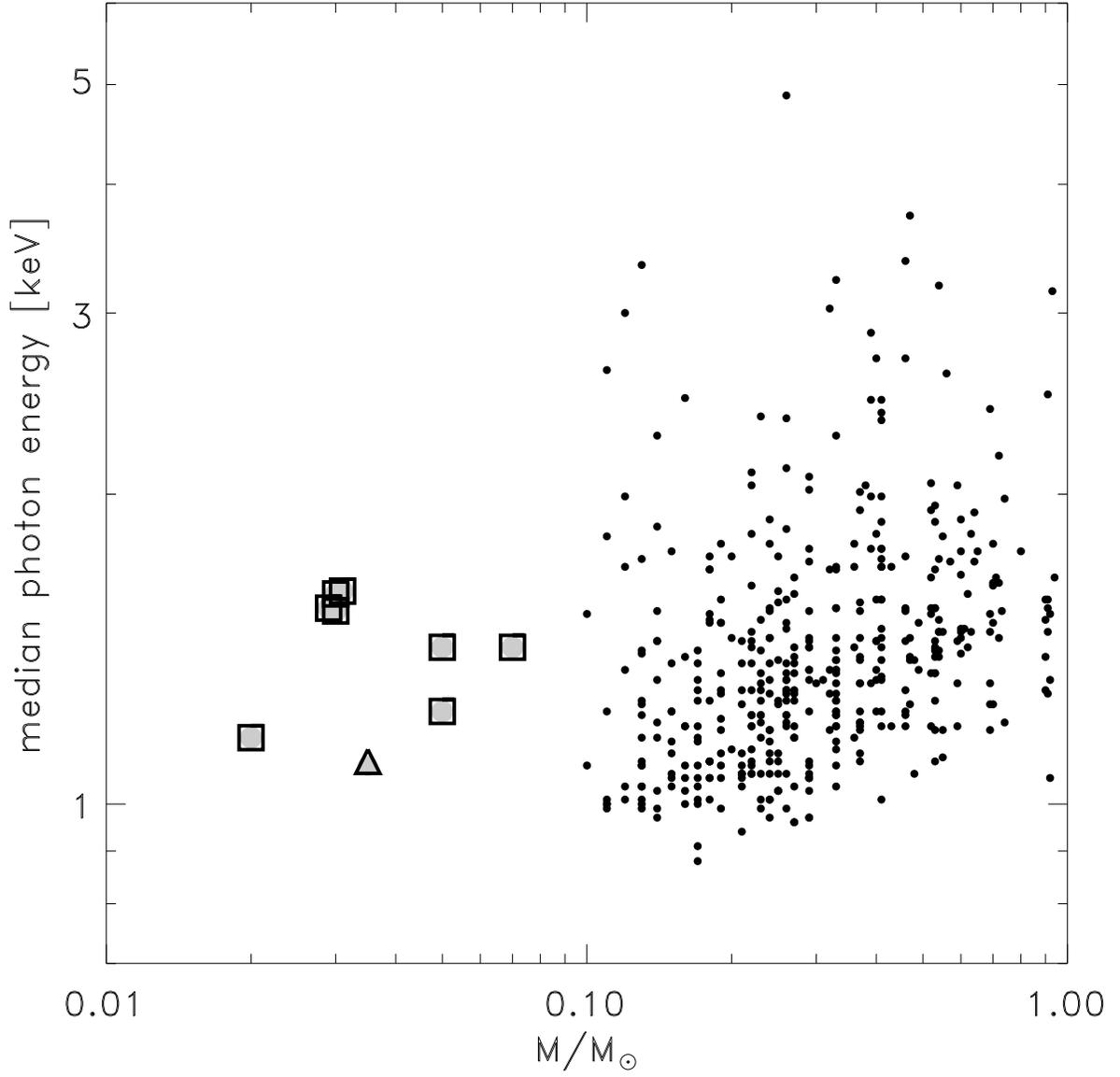}
\caption{Median photon energy versus mass for the COUP-detected BDs (grey
filled squares; the triangle shows COUP\,344 (HC\,722), which was
only detected during a flare) and low-mass stars with low optical extinction
($A_V \leq 5$~mag) from the COUP optical sample
(solid dots; Preibisch et al.\ 2005).
\label{medenergy_mass.fig}}
\end{figure}

The COUP spectra 
of the BDs (with the exception of COUP\,344, which has only 10 net source 
counts) were fitted with single-temperature plasma models; a two-temperature
model was required only for COUP\,280 (HC\,64) to yield an adequate fit to
its spectrum.  Given the rather small numbers ($< 100$) of counts per spectrum,
the spectral fit parameters (plasma temperatures and absorbing hydrogen column  
densities) are subject to relatively large uncertainties.  Rather than 
considering 
the plasma temperatures derived in the fits, we will therefore only consider 
the median energy of the COUP-detected X-ray photons (Table 2), which can 
(for sources with low extinction, $A_{\rm V} \lesssim 5$~mag) be regarded 
as a proxy for the plasma temperature.  Fig.~\ref{medenergy_mass.fig} compares
the median photon energies of the BDs to those of low-mass stars from the 
COUP optical sample (Preibisch et al.\ 2005): the 
median photon energies of the BDs are seen to be generally similar to those 
of the low-mass stars.  The outlier in this plot is the rather 
low median-energy value found for COUP\,344.  Given that this object was 
detected 
only during a flare, it appears that its X-ray emission is 
unusually soft in comparison to that seen from flaring low-mass stars.

\section{X-ray properties of the BDs compared to low-mass ONC stars}
To put the observed X-ray properties of the detected BDs into context, we 
can compare them with those of the X-ray emitting low-mass stars in the ONC\@.
Preibisch et al.\ (2005) define a `COUP optical sample' for
the purpose of investigating relations between the X-ray properties and other 
stellar properties of T\,Tauri stars in the ONC\@. The COUP optical sample 
is a well-defined, homogenous, and representative sample of comprehensively 
characterized young stars, and extends down to objects with spectral types 
of M6.5 with estimated masses around $0.1\,M_\odot$. 
Nearly all objects in the COUP optical sample are safely classified as stars
regardless of the tracks that are used (e.g., have spectral types $\lesssim$ M5).

The SHC04 sample includes stellar-mass sources in addition to the BDs, but
although there is some overlap in mass with the COUP optical sample, it is 
important to note that the two samples cannot be easily quantitatively 
compared for several reasons. First, the COUP optical sample 
is based on a magnitude-limited ($I < 17.5$) sample of ONC stars from 
Hillenbrand (1997), whereas the SHC04 sample consists of mostly much fainter
sources that were selected  as BD candidates from near-infrared 
photometry. The BD candidate selection criteria include a nominal upper
limit brightness of $K \gtrsim 14$, but also some randomly selected brighter
objects. As a consequence, the two samples cannot be considered equally
complete. Second, the SHC04 sample covers a much smaller area than the COUP 
optical sample, thus leading to poorer statistics. Third, SHC04 used the
D'Antona \& Mazzitelli (1997) evolutionary models to estimate masses for
their sample, whereas the masses of the COUP optical sample have been 
estimated using the models of Siess et al.\ (2000). These latter tracks
were adopted as a COUP policy on the basis that they extend across the
full mass range encountered in the ONC stellar population, yielding a more
uniform approach to intercomparing stellar properties (Preibisch et
al.~2005). They do not, however,
extend into the BD mass domain, making it impossible to use them in the
present paper. With these caveats in mind, it is nevertheless instructive 
to compare the properties of the spectroscopic BDs of SHC04 with those of 
the low-mass stars from the COUP optical sample in at least a qualitative way.

\begin{figure}[p]
\plotone{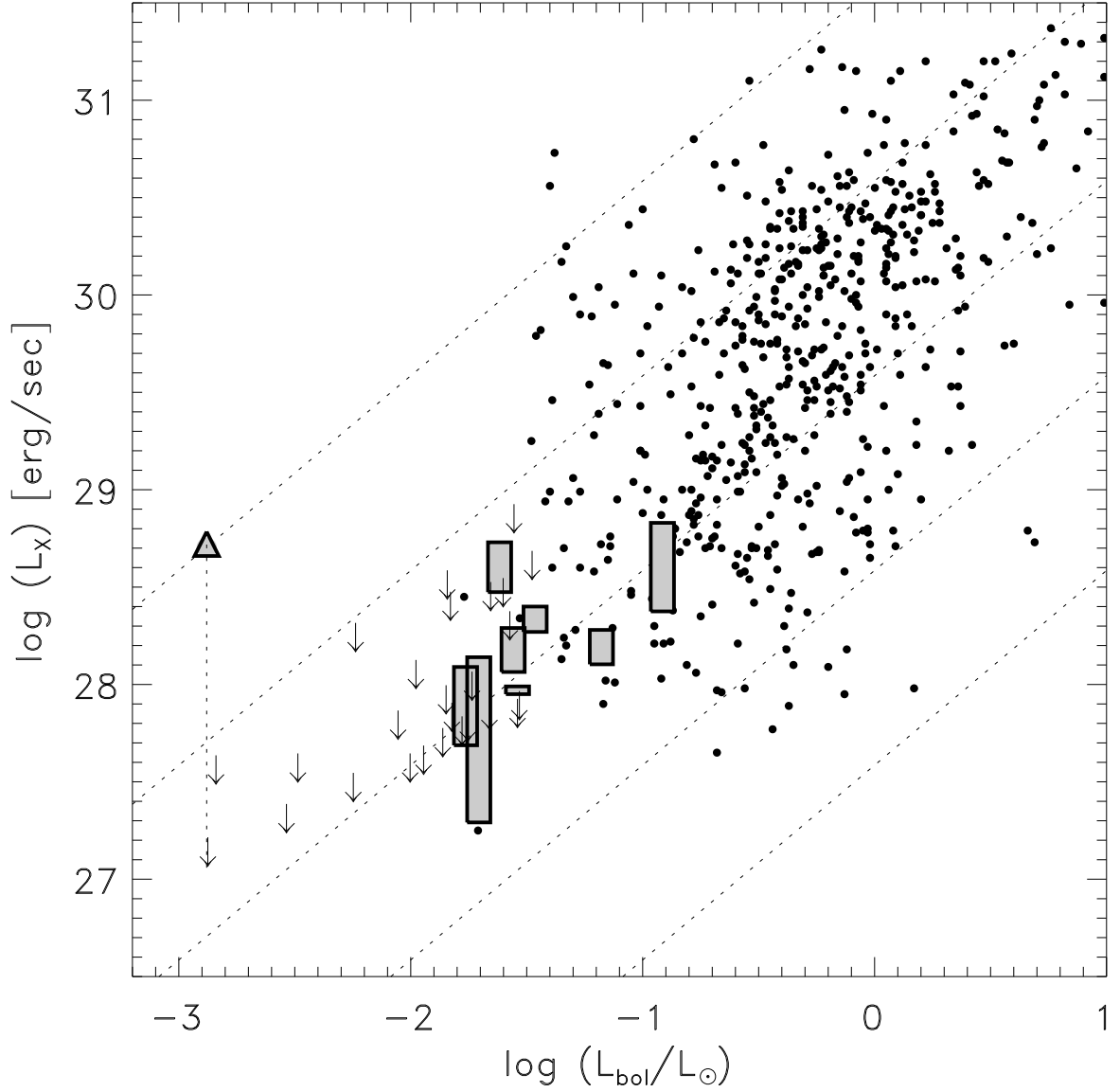}
\caption{X-ray luminosity versus bolometric luminosity for the BDs from
SHC04 (grey filled boxes; arrows for upper limits) and low-mass stars from
the COUP optical sample (solid dots).
The grey filled boxes for the BDs extend from the characteristic X-ray
luminosity found with the MLB analysis (lower edge of the box) to the average
X-ray luminosity (upper edge of the box). For COUP\,344 (HC\,722),
we show the X-ray luminosity during the flare (grey filled triangle) and the
upper limit to the quiescent emission level (arrow).
Some of the
symbols have been slightly shifted along the x-axis to avoid overlaps. The
dotted lines mark $L_{\rm X}/L_{\rm bol}$ ratios of $10^{-2}$, $10^{-3}$,
$10^{-4}$, $10^{-5}$, and $10^{-6}$.
\label{lx_lbol.fig}}
\end{figure}

Figure~\ref{lx_lbol.fig} shows the distribution of X-ray versus bolometric 
luminosities for the BDs and low-mass stars in the ONC\@.  Considering only 
the X-ray detected BDs, the fractional X-ray luminosities lie in the range 
$\log\left(L_{\rm X}/L_{\rm bol}\right) \sim -4$ to $-3$, similar to that
seen for the low-mass stars. Excluding COUP\,344 (HC\,722), which was only 
detected during a large flare, the median fractional X-ray luminosity of the
8 remaining X-ray detected BDs is 
$\log\left(L_{\rm X}/L_{\rm bol}\right) = -3.76$, 
which is identical to the median fractional X-ray luminosity of the X-ray 
detected $0.1-0.25\,M_\odot$ stars in the COUP optical sample. 
Considering the upper limits for undetected BDs as well, we find an upper 
limit for the median 
fractional X-ray luminosity of $\log\left(L_{\rm X}/L_{\rm bol}\right) <-3.8$;
this is only marginally lower than the median value for the detected BDs
because most of the undetected BDs are seen through more extinction
than the X-ray detected BDs (see \S\ref{det_frac.sec}), and therefore most 
upper limits to the extinction-corrected X-ray fluxes are essentially
at the same level as the fluxes for the detected BDs.

\begin{figure}[p]
\plotone{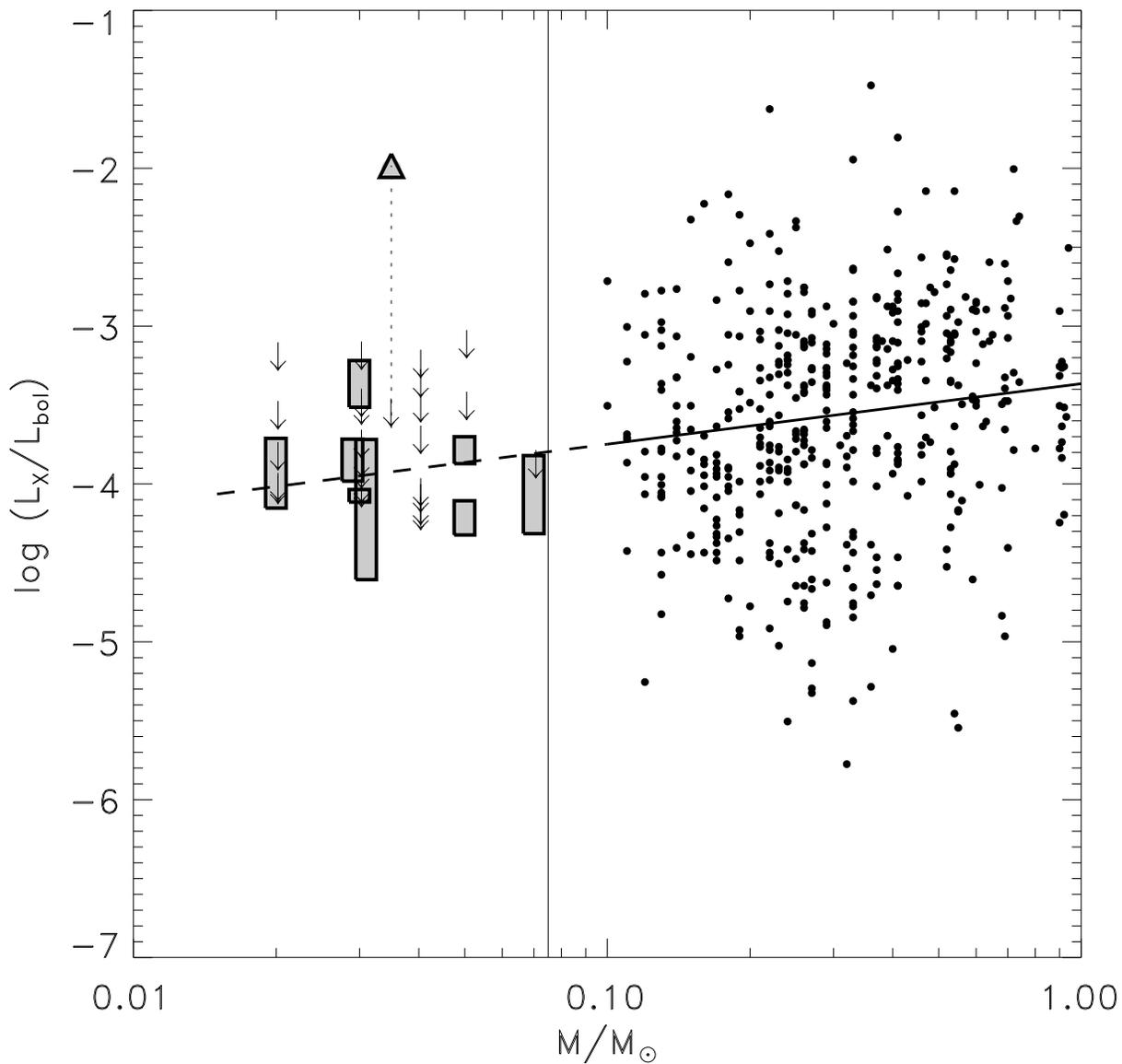}
\caption{Fractional X-ray luminosity
$\log\left(L_{\rm X}/L_{\rm bol}\right)$
versus mass for BDs from the SHC04 sample and low-mass stars from the COUP
optical sample. The symbols are as in Fig.~\ref{lx_lbol.fig}. The solid line
shows a linear regression fit to the low-mass (0.1--2\,$M_\odot$) stars,
while the dashed line shows the same fit extrapolated into the BD regime.
\label{lxlb_mass.fig}}
\end{figure}

Investigating this effect further, Fig.~\ref{lxlb_mass.fig} shows 
the fractional X-ray luminosity ($\log\left(L_{\rm X}/L_{\rm bol}\right)$) 
versus mass for the SHC04 BDs and low-mass stars from the COUP optical sample. 
A regression fit derived in Preibisch et al.\ (2005) for the low-mass 
(0.1--2\,$M_\odot$) stars is plotted, along with an extrapolation of this fit 
into the BD regime.  The BDs seem to follow this extrapolation, in the sense 
that the fractional X-ray luminosity continues to decrease slightly in going 
from low-mass stars to BDs. As discussed in \S\ref{sample.sec}, it is 
possible that the use of near-infrared classifications and a field dwarf 
effective temperature scale by SHC04 have led to a systematic underestimate
of the masses in their sample by a factor of up to 2--3. If so, this would
push them somewhat to the right in Fig.~\ref{lxlb_mass.fig}, but would
not greatly affect the general agreement with the extrapolated trend from
the low-mass stars with optical spectral types. 

\begin{figure}[p]
\plotone{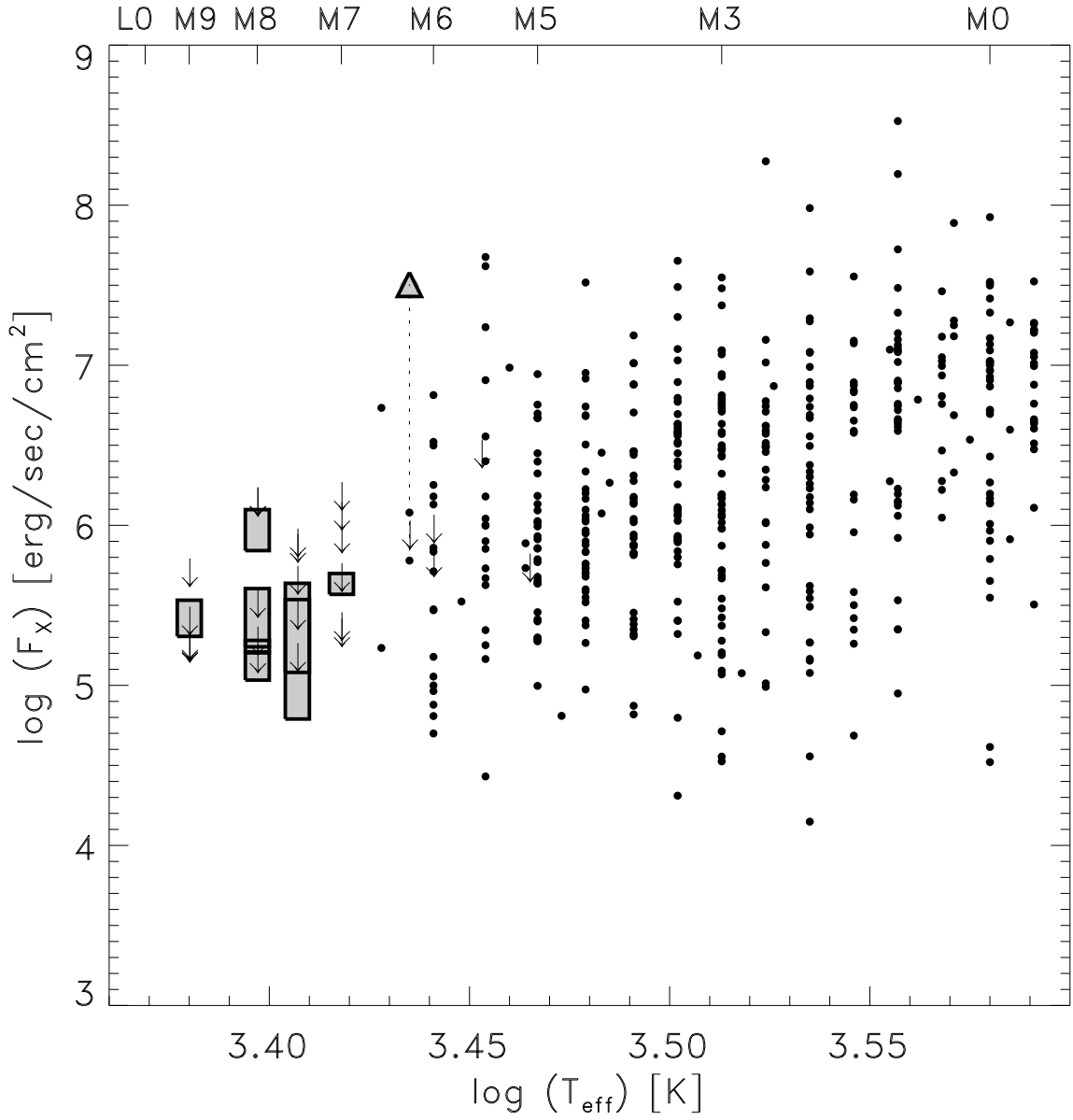}
\caption{X-ray surface flux versus effective temperature for the SHC04
BDs and low-mass stars from the COUP optical sample.
The symbols are as in Fig.~\ref{lx_lbol.fig}.
\label{fx_teff.fig}}
\end{figure}

In Fig.~\ref{fx_teff.fig} we show a plot of the X-ray surface flux (i.e.\ 
the X-ray luminosity divided by the surface area of the object, which is 
computed from its bolometric luminosity and effective temperature) versus 
the effective temperature for the SHC04 BDs and the COUP optical sample of 
low-mass stars. Here the variables are more directly measurable than the 
mass, which may be affected by uncertainties in the theoretical models as
previously discussed. Our data show that the BDs follow the general trend 
of decreasing surface fluxes with decreasing effective temperature. The 
mean surface fluxes in the coronae of the BDs are more than one order of 
magnitude lower than those in early M-type stars, although they are still 
about one order of magnitude higher than the typical average X-ray surface 
flux in the solar corona. 

Again, these deductions would not be significantly affected if the true 
effective temperatures of the BD sample were larger than derived by SHC04. 
An increase in effective temperature by 300\,K would result in a source 
moving by $\sim$\,0.06 dex to the right in Fig.~\ref{fx_teff.fig} and by 
$\sim$\,0.2 dex up, as the surface flux is the X-ray luminosity divided by the 
area of the source, and the area scales as $T_{\rm eff}^{-4}$ for constant 
bolometric luminosity. It can be readily seen in Fig.~\ref{fx_teff.fig}
that the source would still remain well within the extrapolation from the
COUP optical sample. 

\section{Discussion: On the origin of stellar and substellar X-ray emission}
\label{origin.sec}
We now consider the implications of our findings
   on astrophysical concepts underlying the production of magnetic
   activity on BD surfaces.
In solar-type main-sequence stars, the differentiated radiative and convective 
inner structure leads to an $\alpha$--$\Omega$ dynamo, which in 
turn generates magnetic fields which can sustain a hot corona where X-rays
can be emitted. However, both low-mass pre-main sequence stars and BDs are
fully convective, and thus it remains unknown how magnetic fields may arise
in them. Potential alternative dynamo mechanisms are small-scale dynamo 
action in a highly turbulent convection zone (cf.\ Durney et al.~1993; 
Giampapa et al.\ 1996 and 
references therein) or a so-called $\alpha^2$ dynamo, as suggested by K\"uker 
\& R\"udiger (1999). A more detailed discussion can be found in 
Feigelson et al.\ (2003).

\begin{figure}[p]
\plotone{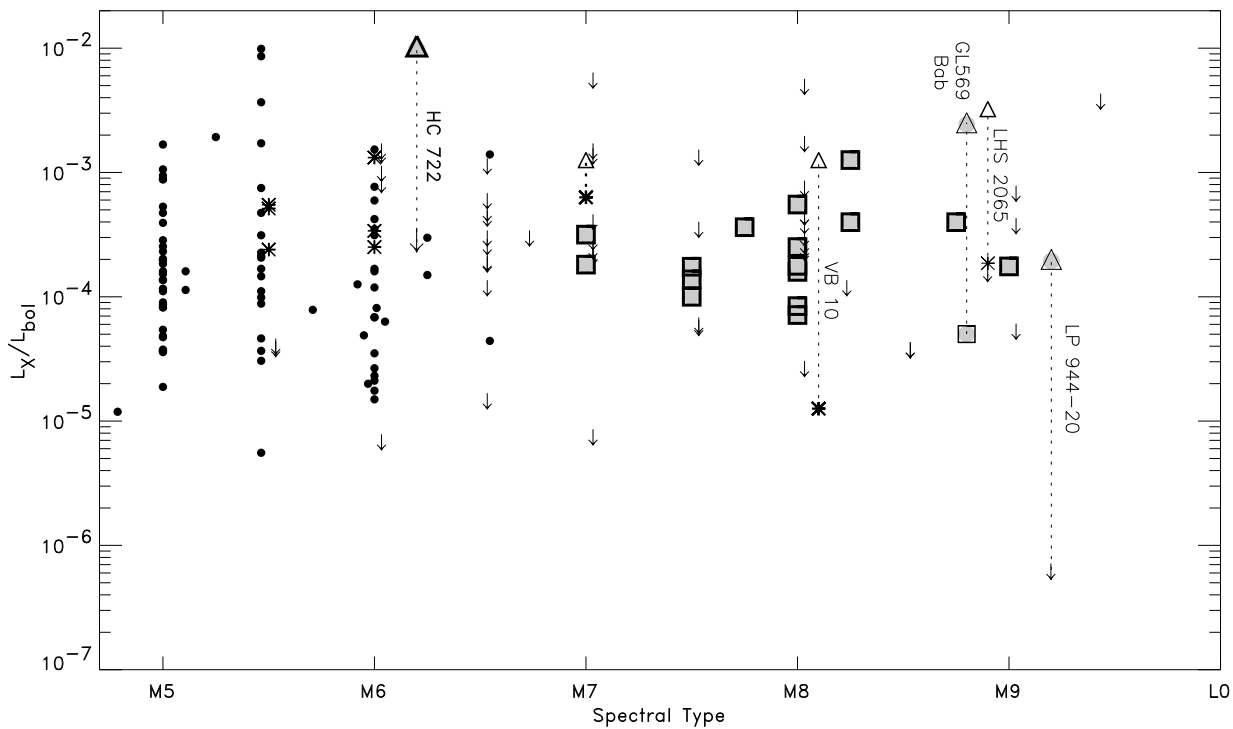}
\caption{
Fractional X-ray luminosity versus spectral type for objects of type M5 or
later. The solid dots show stars in the COUP optical sample.  Data for late
M field stars from Fleming et al.\ (1993) are shown as asterisks.  The BDs
in the ONC from the SHC04 sample and other X-ray detected young BDs 
(as discussed in \S 2) are shown by grey filled squares.  For objects
with strong flares, the values at flare peak are shown by triangles, connected
by dotted lines to the quiescent emission values (or upper limits). Some
individual objects discussed in the text are annotated.  Some symbols have
been slightly shifted along the x-axis to avoid overlaps.
\label{lxlb_spt.fig}}
\end{figure}

Fig.~\ref{lxlb_spt.fig} shows the fractional X-ray luminosities as a function 
of spectral type for the coolest objects in the ONC and compares them to 
equivalent data obtained for other very cool objects.
It confirms that very young ONC BDs
with spectral types in the range 
M6--M9 show essentially the same signatures of activity as older, low-mass 
field stars and BDs with equivalent late-M spectral types. The implication is
that the activity is mainly determined by the effective temperatures of the 
sources and not (so much) by their masses, since the ONC BDs are typically
a factor of four or so lower in mass than field dwarfs of the same spectral
type (Baraffe et al.\ 1998). Similarly, the substantial difference in surface
gravity ($\sim 30$ times higher in an M8 field star compared to an M8 1\,Myr
old BD) appears not to be important. Finally, the main difference between BDs 
and stars, i.e.\ the presence or absence of nuclear hydrogen burning, seems
not to play a role in the X-ray activity. This latter finding is not very
surprising, as it has been known for many years that low-mass pre-main 
sequence stars, which have not yet started nuclear fusion processes, are 
generally strong X-ray sources.

The key to understanding the X-ray activity of  very young BDs seems to 
be that they have relatively early spectral types of M6--M9 for their mass
and are thus still warm enough to maintain a partially-ionized atmosphere. 
As the photospheric properties of these young substellar objects are
essentially the same as those of older stellar objects, it is not surprising 
that their coronal properties are similar to those of low-mass stars. In 
other words, the very young BDs 
do not yet know that they will never undergo hydrogen fusion,
and therefore they behave like low-mass stars. 

While X-ray emission is now clearly established for very cool dwarfs down to 
spectral type M9, no X-ray detections of L- or T-type objects have been 
reported to date, again keeping in mind that COUP\,344/HC\,722 has been
reclassified from L0 to M6--6.5. It is unclear whether the nature of the 
X-ray emission changes at or beyond spectral type $\sim$\,M9, or whether the 
few available X-ray data points on cooler objects are just not sensitive 
enough to provide useful constraints. Observations of very cool field objects 
have provided some indications of changes in the X-ray properties among the 
coolest X-ray detected objects (\S 2). 
The $\sim$\,M9 objects LHS\,2065 (Schmitt \& 
Liefke 2002; note that LHS\,2065 is a very-low mass star and not a BD), 
Gl\,569\,Ba,b (Stelzer 2004), and LP\,944-20 (Rutledge et al.\
2000) showed X-ray flares, but no fully convincing evidence of 
continuous (`quiescent') X-ray
emission was found. In particular, the very restrictive upper limit to 
possible quiescent emission from the field BD LP\,944-20
($\log\left(L_{\rm X}/L_{\rm bol}\right) < -6.3$) found by Mart\'{\i}n 
\& Bouy (2002) supported the idea that the coolest X-ray detected objects may
emit X-rays only during flares, unlike the earlier M dwarfs for which some
kind of quiescent emission has clearly been established.

Our detection of apparently quiescent X-ray emission from the ONC BD HC\,212 
is relevant in this context. Its near-infrared spectral is M9, although again,
it is possible that its optical spectral type would be earlier. Nevertheless,
taken at face value, it would appear to demonstrate that at least young M9 
objects can produce not only X-ray flares, but also more steady X-ray 
emission. This finding is supported by the detection of probably quiescent 
X-ray emission from the young ($\sim$\,10\,Myr) TWA-5B by Tsuboi et al.\ 
(2003), which has an optical spectral type of M8.5--9 (Neuh\"auser et al.\ 
2000). Thus it appears as though quiescent emission may indeed by possible
at a spectral type of M9 when the source is young, although it remains 
unclear whether this is only because young M9 sources may be significantly
warmer than their field star counterparts, as discussed above in 
\S\ref{sample.sec}.

Another important result in this context is the observed drop in the typical 
X-ray surface fluxes by about one order of magnitude between early and late~M 
spectral types (Fig.~\ref{fx_teff.fig}); this suggests that the coronal 
properties of very cool stars do indeed change over the temperature range
$\sim 3700$\,K to $\sim 2400$\,K\@. Mohanty \& Basri (2003) noted an analogous
change in the chromospheric properties, as traced by $F_{{\rm H}\alpha}$ and
$L_{{\rm H}\alpha}/L_{\rm bol}$, over a similar spectral type range (M4--M9)
in field stars.

Finally, we note that strong changes in the coronal properties are expected 
across the M- to L-type transition from observational indications as well
as from theoretical considerations. Observational evidence comes from the 
sharp and strong  drop in H$\alpha$ emission (a tracer of chromospheric
activity) around spectral type M9--L0 for field dwarfs (Gizis et al.\ 2000;
Mohanty \& Basri 2003). The lack of chromospheric activity in the ultra-cool 
L-type dwarfs is most likely related to the fact that the atmospheres of 
objects cooler than $T_{\rm eff} \sim 2400$\,K (corresponding to spectral 
type M9) are essentially neutral and have a very high electrical resistivity, 
in which the rapid decay of currents prevents the buildup of magnetic free 
energy and therefore cannot provide support for magnetically-heated 
chromospheres and coronae (see Fleming et al.\ 2000; Mohanty et al.\ 2002). 
These ultra-cool dwarfs should therefore not produce the same kind of 
quiescent X-ray emission as originates from magnetically-confined plasma in 
the coronae of late-type stars. However, it is interesting to note that 
several studies have found flaring H$\alpha$ emission in some L- and T-type 
dwarfs (e.g., Burgasser et al.\ 2000; Hall 2002; Liebert et al.\ 2003). These 
discoveries suggest that even the ultra-cool dwarfs can show some kind of 
magnetic activity, although it probably has a different nature to that seen 
in the hotter M-type dwarfs. One possible explanation is that rapidly-rising 
individual flux tubes from the interior of these objects dissipate currents 
in the atmosphere and cause flares (Mohanty et al.\ 2002). 

\section{Conclusions}
The data of the {\it Chandra\/} Orion Ultradeep Project have provided X-ray 
detections for 9 of the 34 spectroscopically-confirmed BDs of SHC04 in the 
central part of the Orion Nebula Cluster. The low detection fraction is not 
only related to the intrinsic faintness of these objects, but also to the 
foreground extinction which is, in many cases, substantial. 
Considering only the BDs with $A_V \leq 5$\,mag, the detection fraction 
is 7/16 = 44\%.
For all but one
of the X-ray detected late M-type BDs, an analysis of their X-ray lightcurves 
revealed evidence for continuous (`quiescent') emission in addition to several 
large flares. Our results extend the spectral type of the coolest known object 
with clear evidence for quiescent X-ray emission down to M9, although we note
that corresponding spectral type was obtained via near-infrared spectroscopy
and should be confirmed via optical spectroscopy if possible.

Our results show no evidence that BDs with ages around
   1~Myr are significantly less magnetically active than late-type
   stars of similar ages.  A gradual trend of decreasing fractional
   X-ray luminosity or X-ray surface flux is seen as one progresses
   from 1~M$_\odot$ to 0.1~M$_\odot$ to BD masses, but no sharp change
   in X-ray properties at the substellar boundary is seen.  The X-rays
   are seen in both flare and apparently quiescent modes.  These results are
   consistent with previous $Chandra$ studies of young BDs in nearby
   star-forming regions, although these involved smaller samples and
   shorter exposures. 

A comparison of the X-ray properties of the late M-type BDs with those of 
late-type stars suggests that they share a common X-ray emission mechanism 
which is governed primarily by the photospheric temperature, not by the mass 
or surface gravity of the source. The absence of any clear transition in 
X-ray properties across the stellar/substellar mass boundary implies that
this conclusion is robust against uncertainties in the detailed classification 
of sources on either side of that boundary.

%%%%%%%%%%%%%%%%%%%%%%%%%%%%%%%%%%%%%%%%%%%%%%%%%%%%%%%

\acknowledgements
COUP is supported by {\it Chandra\/} Guest Observer grant SAO GO3-4009A
(E. Feigelson, PI). Further support was provided by
the Chandra ACIS Team contract NAS8-38252.
BS, EF, GM, and SS acknowledge financial support from an 
Italian MIUR PRIN program and from INAF\@. We would like to thank
the referee, Subhanjoy Mohanty, for his detailed, insightful, and
helpful comments on this paper. 

%%%%%%%%%%%%%%%%%%%%%%%%%%%%%%%%%%%%%%%%%%%%%%%%%%%%%%%%%%%%%%%%%%%%

Facility: CXO(ACIS)

\bibliography{aj-jour}

\clearpage

%%%%%%%%%%%%%%%%%%%%%%%%%%%%%%%%%%%%%%%%%%%%%%%%%%%%%%%%%%%%%%%%%%%%%%%%%

\begin{deluxetable}{rrrrrrrrrrrrr} %\hline
\tablecolumns{13}
\tabletypesize{\small}
\tablewidth{0pt}
\tablecaption{X-ray detected brown dwarfs in the ONC: Near-infrared properties\label{NIRprop.tab}}
\tablehead{
\colhead{COUP} && \multicolumn{5}{c}{VLT} && \multicolumn{5}{c}{Keck} \\
\cline{3-7} \cline{9-13}
\colhead{\#} && \colhead{\#} & \colhead{$\theta$} &
\colhead{$J$} & \colhead{$H$} & \colhead{$K_s$} && \colhead{HC \#} &
\colhead{$A_V$} & \colhead{SpT} &
\colhead{$M$} & \colhead{$\log(L_{\rm bol})$} \\
& &&  (\arcsec) & \multicolumn{3}{c}{(mag)} && &
\colhead{(mag)} && \colhead{($M_\odot$)} &
\colhead{($L_\odot$)} }
\startdata
 280 &&  133 & 0.29 & 16.27 & 15.24 & 14.50 &&  64 & 4.46 & M7-9 & 0.03 &$-1.617$\\
 344 &&  201 & 0.27 & 17.10 & 16.58 & 16.12 && 722 & 0.00 & M6-6.5&0.03 &$-2.880$\\
 371 &&  225 & 0.25 & 15.06 & 13.90 & 13.04 &&  90 & 6.77 & M7.5 & 0.07 &$-0.915$\\
 764 &&  558 & 0.10 & 16.25 & 15.57 & 14.91 && 565 & 5.60 & M8   & 0.02 &$-1.764$\\
 941 &&  724 & 0.03 & 16.35 & 15.45 & 14.75 && 594 & 2.20 & M7.5 & 0.03 &$-1.707$\\
1125 &&  880 & 0.14 & 14.53 & 13.72 & 13.22 &&  22 & 1.78 & M8   & 0.05 &$-1.178$\\
1194 &&  934 & 0.16 & 16.58 & 15.12 & 14.38 &&  55 & 1.90 & M8   & 0.03 &$-1.540$\\
1255 &&  999 & 0.10 & 15.58 & 14.82 & 14.17 && 212 & 3.41 & M9   & 0.03 &$-1.559$\\
1313\tablenotemark{a} && 1055 & 0.14 & 15.14 & 14.45 & 13.94 && 114 & 1.44 & M7   & 0.05 &$-1.464$\\
\enddata
\tablecomments{
Columns 1--6 are from Getman et al.\ (2005) with near-infrared data in the
2MASS photometric system from the VLT unified catalog of the central 
$7\arcmin \times 7\arcmin$ by McCaughrean et al.\ (2005, in preparation). 
$\theta$ is the angular offset between the COUP and VLT positions for each 
source. \\
Columns 7--11 are from SHC04.
}
\tablenotetext{a}{COUP\,1313 has an optical counterpart in Herbst et 
al.\ (2002), \#10626 with $<V> = 18.48$ and variability range $\Delta V = 0.39$.}
\end{deluxetable}

%%%%%%%%%%%%%%%%%%%%%%%%%%%%%%%%%%%%%%%%%%%%%%%%%%%%%%%%%%%%%%%%%%%%%%%%%%%

\begin{deluxetable}{rrrrrrrrrr} % \hline
\tablecolumns{10}
\tabletypesize{\small}
\tablewidth{0pt}
\tablecaption{X-ray detected brown dwarfs in the ONC: X-ray properties\label{Xrayprop.tab}}
\tablehead{
\colhead{COUP} & \colhead{IAU} & \colhead{NetCts} &
\colhead{$\log P_{KS}$} & \colhead{\#BB} & \colhead{Max/} &
\colhead{$<E>$} & \colhead{$\log L_t$} & \colhead{$\log L_{t,c}$} &
\colhead{$\log$} \\
&&&&&Min& (keV) & \multicolumn{2}{c}{erg/sec} & $L_{\rm X}/L_{\rm bol}$ 
  }
\startdata
 280 & 053507.0-052500 & 49 &$-0.7$& 1 &  1 & 1.2 &  28.2  & 28.7 & $
-3.3$\\
 344\,f \tablenotemark{a}& 053509.7-052406 &  8 &$-1.0$& 1 &  1 & 1.1 &  28.7 & 28.7 & $-2.0 $\\ % flare
 344\,q \tablenotemark{a}& 053509.7-052406 & $<5$ &\nodata&\nodata&\nodata&\nodata& $<27.1$& $<27.1$ &$<-3.6$\\ % quiescent
 371 & 053510.3-052451 & 63 &$-0.9$& 1 &  1 &  1.4 &  27.6  & 28.8 & $
-3.9$\\
 764 & 053516.0-052153 & 65 &$-4.0$& 3 & 22 &  1.2 &  28.1  & 28.1 & $
-3.8$\\
 941 & 053518.0-052141 & 43 &$-4.0$& 2 &  8 &  1.6 &  28.0  & 28.1 & $
-3.8$\\
1125\tablenotemark{b}& 053520.9-052534 & 26 &$-1.0$& 1 &  1 &  1.4 &  28.2  & 28.3 & $
-4.2$\\
1194 & 053522.1-052507 & 24 &$-0.4$& 1 &  1 &  1.6 &  27.9  & 28.0 & $
-4.1$\\
1255 & 053523.5-052350 & 46 &$-3.7$& 2 &  3 &  1.6 &  28.1  & 28.3 & $
-3.8$\\
1313 & 053525.0-052438 & 94 &$-2.0$& 3 &  7 &  1.2 &  28.3  & 28.4 & $
-3.7$\\ 
\enddata

\tablecomments{%
Columns 1--6 and 8--9 are from Getman et al.\ (2005), while Columns 7 and 
10 are calculated here. \\
Column~3: net (i.e.\ background-subtracted) number of X-ray photons detected \\
Column~4: Logarithm of the nonparametric one-sample Kolmogorov-Smirnov (KS) 
test significance for the null hypothesis of a constant source \\
Column~5: Number of segments into which the lightcurve 
was segmented by the Bayesian Block (BB) parametric model\\
Column~6: Ratio of the count rates in the highest and lowest segment\\
Column~7: Median energy of the detected photons for each source \\
Column~8: Observed X-ray luminosity integrated over the 
[$0.5 - 8.0$]\,keV band \\
Column~9: Extinction-corrected [$0.5 - 8.0$]\,keV band luminosity \\
Column~10: Ratio of total X-ray luminosity to stellar bolometric luminosity 
}
\tablenotetext{a}{COUP\,344 (HC\,722) has two entries in this table:
344\,f lists the X-ray properties during the flare, while 344\,q lists
the upper limits for the periods outside the flare.}
\tablenotetext{b}{COUP\,1125 lies on an ACIS chip gap with an 
effective exposure of only 364\,ksec.}
\end{deluxetable}

%%%%%%%%%%%%%%%%%%%%%%%%%%%%%%%%%%%%%%%%%%%%%%%%%%%%%%%%%%%%%%%%%%%%%%%%%%%

\begin{deluxetable}{rrrrr} % \hline
\tablecolumns{5}
\tabletypesize{\small}
\tablewidth{0pt}
\tablecaption{X-ray detected brown dwarfs in the ONC: MLB 
analysis of variability\label{countrates.tab}}
\tablehead{
\colhead{COUP} & \colhead{HC} & \colhead{average} &
\colhead{charact.} & \colhead{\# of}\\
&&cts ksec$^{-1}$ & cts ksec$^{-1}$ & flares 
  }
\startdata
 280 &  64 & 0.058 &  0.035 & 2\\
 344 & 722 & 0.012 &$<0.006$ & 1 \\
 371 &  90 & 0.074 &  0.028 & 2 \\
 764 & 565 & 0.077 &  0.033 & 2\\
 941\tablenotemark{a}& 594 & 0.051 &  0.008 & 2\\
1125 &  22 & 0.031 &  0.022 & 1\\
1194 &  55 & 0.029 &  0.029 & 0\\
1255 & 212 & 0.054 &  0.035 & 1\\
1313 & 114 & 0.111 &  0.090 & 1\\
\hline
\enddata

\tablenotetext{a}{In COUP\,941, the characteristic count rate is below
the 3$\sigma$ limit.}
\end{deluxetable}

%%%%%%%%%%%%%%%%%%%%%%%%%%%%%%%%%%%%%%%%%%%%%%%%%%%%%%%%%%%%%%%%%%%%%%%%%%%

\begin{deluxetable}{rrrrrrrrr}
\centering
\tabletypesize{\small}
\tablewidth{0pt}
\tablecolumns{9}
\tablecaption{Upper limits for the COUP undetected BDs from SHC04.  
\label{Slesnick_UL.tab}}
\tablehead{
  \colhead{HC00} &  \colhead{LC} &  \colhead{Exp} &
  \colhead{$A_V$} &  \colhead{CF} &  \colhead{CF$_c$} &
  \colhead{$\log L_t$} &  \colhead{$\log L_{t,c}$} & 
  \colhead{$\log (L_{\rm X}/L_{\rm bol})$}
}
\startdata
20  &  6 &  424 &  3.77 &  0.65 &  1.9 &  $<$27.3 &  $<$27.8 & $<-4.3$ \\
62  &  4 &  837 &  5.85 &  0.71 &  2.9 &  $<$26.9 &  $<$27.5 & $<-4.1$ \\
70  &  6 &  641 &  9.48 &  0.79 &  4.1 &  $<$27.3 &  $<$28.0 & $<-3.7$ \\
111 &  4 &  838 &  8.61 &  0.77 &  3.6 &  $<$27.0 &  $<$27.6 & $<-4.1$ \\
123 &  6 &  834 &  7.89 &  0.75 &  3.4 &  $<$27.1 &  $<$27.8 & $<-4.3$ \\
167 & 20 &  823 &  9.19 &  0.79 &  4.0 &  $<$27.7 &  $<$28.4 & $<-3.6$ \\
210 &  9 &  834 &  4.81 &  0.66 &  2.2 &  $<$27.2 &  $<$27.8 & $<-4.2$ \\
221 & 16 &  331 &  4.43 &  0.65 &  2.1 &  $<$27.9 &  $<$28.4 & $<-3.6$ \\
365 & 18 &  797 &  8.23 &  0.76 &  3.8 &  $<$27.6 &  $<$28.3 & $<-3.5$ \\
372 & 25 &  831 &  2.04 &  0.60 &  1.1 &  $<$27.6 &  $<$27.9 & $<-4.0$ \\
400 &  9 &  777 &  2.16 &  0.60 &  1.1 &  $<$27.2 &  $<$27.5 & $<-4.1$ \\
403 &  5 &  733 &  6.78 &  0.71 &  2.9 &  $<$27.1 &  $<$27.7 & $<-4.2$ \\
429 &  8 &  825 &  5.53 &  0.68 &  2.4 &  $<$27.2 &  $<$27.8 & $<-4.0$ \\
433 &  5 &  770 & 10.44 &  0.81 &  4.4 &  $<$27.1 &  $<$27.8 & $<-3.9$ \\
515 &  9 &  834 & 23.29 &  1.20 & 13.0 &  $<$27.5 &  $<$28.5 & $<-3.6$ \\
529 & 14 &  833 & 25.18 &  1.20 & 14.3 &  $<$27.7 &  $<$28.8 & $<-3.3$ \\
559 & 28 &  831 &  0.00 &  0.61 &  0.6 &  $<$27.7 &  $<$27.7 & $<-4.2$ \\
709 &  4 &  431 &  2.11 &  0.60 &  1.1 &  $<$27.1 &  $<$27.4 & $<-4.0$ \\
724 &  9 &  779 &  0.00 &  0.61 &  0.6 &  $<$27.2 &  $<$27.2 & $<-3.8$ \\
725 &  4 &  753 &  9.40 &  0.79 &  4.0 &  $<$27.0 &  $<$27.7 & $<-3.8$ \\
728 &  8 &  365 &  5.91 &  0.70 &  2.7 &  $<$27.6 &  $<$28.2 & $<-3.2$ \\
729 & 13 &  821 & 14.73 &  0.96 &  7.0 &  $<$27.6 &  $<$28.4 & $<-3.3$ \\
743 &  4 &  785 &  5.61 &  0.68 &  2.5 &  $<$26.9 &  $<$27.5 & $<-3.6$ \\
749 &  7 &  785 &  3.10 &  0.60 &  1.4 &  $<$27.1 &  $<$27.5 & $<-3.3$ \\
764 &  7 &  316 &  7.06 &  0.73 &  3.1 &  $<$27.6 &  $<$28.2 & $<-3.8$ \\
\enddata
\tablecomments{~ \\
Col.\ 1: Source number from Hillenbrand \& Carpenter (2000) \\
Col.\ 2: Limiting number of counts from COUP image (see text) \\
Col.\ 3: COUP effective exposure time in ksec \\
Col.\ 4: Visual absorption from SHC04 \\
Col.\ 5: Conversion factor from count rate to flux in units of
         $10^{-14}$ erg s$^{-1}$ cm$^{-2}$ (ct ks$^{-1}$)$^{-1}$ as
         observed (see text) \\
Col.\ 6: Conversion factor corrected for absorption \\
Col.\ 7: Total $[0.5 - 8]$\,keV band luminosity as observed \\
Col.\ 8: Total band luminosity corrected for absorption \\
Col.\ 9: Upper limit on $\log (L_{\rm X}/L_{\rm bol})$
}
\end{deluxetable}

\end{document}